\documentclass[aps,twocolumn,nofootinbib,prd,eqsecnum,showpacs,showkeys,preprintnumbers]{revtex4-2}
\usepackage{graphicx}
\usepackage{epstopdf}
\usepackage{wrapfig}
\usepackage{caption}
\usepackage{subcaption}
\usepackage{amsmath}
\usepackage{amsfonts}
\usepackage{amssymb}
\usepackage{xcolor}
\usepackage{bm}
\usepackage{natbib}
\usepackage{epstopdf}
\usepackage{url}
\usepackage{footnote}
\usepackage{textcomp}
\usepackage{ulem}
\usepackage{esint}
\usepackage[unicode=true, pdfusetitle,
 bookmarks=true,bookmarksnumbered=false,bookmarksopen=false,
 breaklinks=false,pdfborder={0 0 1},backref=false,colorlinks=false]{hyperref}
\usepackage{multirow}
\usepackage{pifont}
\usepackage{lipsum}
\usepackage[mathlines]{lineno}
\usepackage[newcommands]{ragged2e}
\begin{document}
\title{Higgs inflation model with non-minimal coupling in hybrid Palatini approach}
\author{Brahim Asfour}
\email{brahim.asfour@ump.ac.ma}
\author{Aatifa Bargach}
\email{a.bargach@ump.ac.ma}
\author{Ahmed Errahmani}
\email{ahmederrahmani1@yahoo.fr}
\author{Taoufik Ouali}
\email{t.ouali@ump.ac.ma}
\affiliation{Laboratory of Physics of Matter and Radiation, \\
University of Mohammed first, BP 717, Oujda, Morocco}

\date{\today }

\begin{abstract}
In this paper, we construct a hybrid metric Palatini approach in which the Palatini scalar curvature is non minimally coupled to  the scalar field. We derive the Einstein's field equations, the equations of motion of the scalar field. Furthermore, the background and the perturbative parameters are obtained by means of Friedmann equations in the slow roll regime. The analysis of cosmological perturbations allowed us to obtain the main inflationary parameters such as the scalar spectral index $n_s$ and the tensor to scalar ratio $r$. In this perspective, as an application of our analysis, we consider the Higgs field with quartic potential which plays the inflaton role, and we show that predictions of Higgs hybrid inflation are in good agreement with the recent observational data \cite{Akrami:2018odb}.
\end{abstract}
\maketitle
\newpage
\section{INTRODUCTION}

 One of the most successful approach to explain the early Universe phenomena is the cosmic inflation  \cite{1, Albrecht, Starobinsky, Linde2, Sato, Linde1}, i.e.  the accelerated expansion of the early Universe. This important idea has the fundamental implication that the shortcomings of the standard cosmology could be explained in an elegant way and also the origin of anisotropies observed in the cosmic microwave background (CMB) radiation itself becomes a natural theory \cite{6, 7, 8, 9, 10,Bennett, 11, 12, 2, Larson}. In this context, one of the most remarkable evolution in modern physics was the observational constraints that ruled out many inflationary models if they are not supported by the observational data \cite{Akrami:2018odb,BICEP:2021xfz,AtacamaCosmologyTelescope:2013swu,WMAP:2012fli}. Indeed, the observational value of the spectral index $n_{s}$ and the analysis of the consistent behavior of this spectral index versus the tensor to scalar ratio $r$, help to reduce the number of these inflationary models. In fact, recent observational data \cite{Akrami:2018odb} imposes  constraints on both parameters: an upper limit on the tensor to scalar ratio $r < 0.1$ (Planck alone) at a 95\% confidence level (CL) as well as a value of the spectral index $n_{s}=0.9649\pm 0.0042$ quoted at $68$\% CL.\par
The most famous illustration of the scenario of inflation is that the Higgs boson of the standard model acts as the inflaton \cite{bargach2020nonminimal, cervantes1995induced, Bezrukov:2007ep, rehman2010higgs, bezrukov2019some, raatikainen2019higgs}. There are two approaches to obtain the field equations from the Lagrangian of this theory, known as the metric formalism and the Palatini formalism. In the original scenario \cite{Bezrukov:2007ep}, general relativity is based on the metric formulation where all gravitational degrees of freedom are carried by the metric field and the connection is fixed to be the Levi-Civita one. However, in the Palatini formulation of gravity, the metric and the connection are two independent variables. It seems important enough to mention that both formulations lead to the usual Einstein field equations of motion in minimally coupled scenarios. However, in the non-minimal coupling (NMC), different approaches lead to different predictions even when the Lagrangian density of the theory has the same form. In addition, the assumption of considering a non-minimal coupling to gravity is important to sufficiently flatten the Higgs potential at large field values \cite{Bezrukov:2007ep} in order to be in concordance with observations. A remarkable difference between metric and Palatini formalism arise from observational consequences. Indeed, predictions of Palatini Higgs inflation give an extremely small tensor to scalar ratio \cite{Rasanen:2017ivk, Enckell:2018kkc} compared to the metric formalism.  Another interesting feature of Palatini Higgs inflation is that it has a higher cutoff scale, above which the perturbation theory breaks down, than the metric theory \cite{Bauer:2010jg}. For reviews on this topic, please see Ref. \cite{Rubio:2018ogq} for the metric and Ref. \cite{Tenkanen:2020dge} for the Palatini Higgs inflation. Furthermore, Palatini Higgs inflation lowers the spectral index for the primordial spectrum of density perturbations, and reduces  the required number of e-folds to answer hot big bang cosmology puzzles \cite{Rubio:2019ypq}. In this paper, we construct an alternative way to connect the metric and the Palatini Higgs inflation dubbed as hybrid metric Palatini Higgs inflation\footnote {Hybrid metric Palatini Higgs inflation was also considered in \cite{He:2022xef} using the Einstein-frame analysis. The framework considered by the authors of this paper, however, is completely based on a non minimal coupling between the Higgs field and both the metric and the Palatini Ricci scalar curvature. Whereas, in our analysis,  non minimal coupling is between the Higgs field and the Palatini scalar curvature alone.}. The idea of hybrid metric Palatini scenario was already studied in \cite{Harko:2011nh}, where an $f(R)$ Palatini correction to the Einstein-Hilbert Lagrangian was added. This kind of hybrid theory typically develops when perturbative quantization techniques are taken into account on Palatini formalisms \cite{Flanagan:2003iw} and it is connected to non-perturbative quantum geometries in interesting ways \cite{Olmo:2008nf}. Moreover, using  the scalar-tensor representation of a metric Palatini formalism was found to be useful in cosmology with respect to local experiments and overcoming any  matter instabilities if the scalar field is only weakly connected to matter. In this regard, wormhole geometries, cosmological and astrophysical applications have been examined in \cite{Capozziello:2012ny}, where it has been shown that accelerating solutions are possible. Dynamical system in hybrid metric Palatini context is also analyzed in \cite{Tamanini:2013ltp}.
\par

In the present paper, we consider a novel approach to modified gravity, in which one combines elements from both theories \cite{capozziello2015hybrid}. Thus, one can avoid shortcomings that emerge in pure metric or Palatini approaches, such as the cosmic expansion and the structure formation. This recent formalism is called hybrid metric-Palatini gravity, and it consists to add a Palatini scalar curvature to the Einstein-Hilbert action. The benefit of this kind of hybrid metric Palatini is to preserve the advantage of the minimal  metric approach and improving the non minimal coupling from the metric by the Palatini one.

The aim of this work is to study the non-minimally coupled Higgs inflation under the hybrid metric-Palatini approach and check the results in light of the observational data \cite{Akrami:2018odb}. 

The paper is structured as follows. In Section \ref{secII}, from the action, we derive the basic field equations of the inflation model with NMC in a hybrid metric Palatini formalism. In Section \ref{secIII}, we present the Friedmann equation and we apply the slow roll conditions on it. In Sections \ref{secIV} and \ref{secV}, we analyze cosmological perturbations. In Section \ref{secVI}, we consider a Higgs inflation model and we check its viability. Finally, we summarize and conclude in section \ref{secVII}.
\section{SETUP}
\label{secII}
We consider a hybrid Palatini model where the scalar field is non-minimally coupled to the gravity. Its action is described by
	\begin{eqnarray}
\nonumber S=\int d^4x \sqrt{-g}\left( \frac{M_p^2}{2}R + \frac{1}{2} \xi \phi^2 \hat{R} +\mathcal{L_{\phi}}(g_{\mu \nu},\phi)\right), \label{eq1}\\
\end{eqnarray}
where $g$ is the determinant of the metric tensor $g_{\mu \nu}$, $M_p$ is the Planck mass, R is the Einstein-Hilbert curvature term, determined by the metric tensor $g_{\mu \nu}$, and $\hat{R}$ is the Palatini curvature, depending on the metric tensor $g_{\mu \nu}$  and on the connection $\Gamma_{\beta \gamma}^{\alpha}$ which is considered as an independent variable $\hat{R}=\hat{R}(g_{\mu \nu},\Gamma_{\beta \gamma}^{\alpha})$ \cite{24}. $\xi$ is the coupling constant, and $\mathcal{L_{\phi}}$  the lagrangian density of the scalar field $\phi$, which takes the following form 
\begin{eqnarray}
\mathcal{L_{\phi}}=-\frac{1}{2} \nabla_{\mu} \phi \nabla^{\mu} \phi - V(\phi),\label{eq2}
\end{eqnarray}
where $V(\phi)$  is the scalar field potential.\par
The variation of this action with respect to the independent connection gives 
\begin{eqnarray}
\nabla_{\sigma}(\xi \phi^2\sqrt{-g}g^{\mu \nu})=0.
\end{eqnarray}
The solution of this equation reveals that the independent connection is the Levi- Civita connection of
the conformal metric $\hat{g}_{\mu \nu}=\xi \phi^2g_{\mu \nu}$,
 
\begin{eqnarray}
\nonumber \hat{\Gamma}^{\rho}_{ \mu \sigma}&=&\frac{1}{2}\hat{g}^{\lambda \rho}\left( \partial_\mu \hat{g}_{\lambda \sigma}+\partial_\sigma  \hat{g}_{\mu\lambda }-\partial_\lambda \hat{g}_{\mu \sigma}\right) \\
\nonumber &=&\overset{}{{\Gamma}}{^{\rho}_{ \mu \sigma}}+\frac{\omega}{\phi}(\delta^\rho_\sigma\partial_\mu(\phi)+\delta^\rho_\mu \partial_\sigma(\phi)-{g}_{\mu \sigma}\partial^\rho( \phi) ),\\
\label{eq.2.4}
\end{eqnarray}
with $\omega = 1$ corresponds to the Palatini approach and $\omega = 0$ to
the metric one. The curvature tensor $\hat{R}_{\mu \nu}$ is given in terms of the independent connection $\hat{\Gamma}^{\alpha}_{\beta \gamma}$ \cite{24}
\begin{equation}
\hat{R}_{\mu \nu}=\hat{\Gamma}^{\alpha}_{\mu \nu,\alpha }-\hat{\Gamma}^{\alpha}_{\mu \alpha,\nu }+\hat{\Gamma}^{\alpha}_{\alpha\lambda}\hat{\Gamma}^{\lambda}_{\mu \nu}-\hat{\Gamma}^{\alpha}_{\mu\lambda}\hat{\Gamma}^{\lambda}_{\alpha \nu}, \label{eq.2.5}
\end{equation}
by using Eq.$\eqref{eq.2.4}$, we can rewrite Eq.$\eqref{eq.2.5}$ as
\begin{widetext}
\begin{eqnarray}
\nonumber \hat{R}_{\mu \nu}={R}_{\mu \nu}+\frac{\omega}{\phi^2}\left[ 4\nabla_{\mu}\phi \nabla_{\nu}\phi-{g}_{\mu \nu}(\nabla \phi)^2-2\phi\left( \nabla_{\mu}\nabla_{\nu}+\frac{1}{2}{g}_{\mu \nu}\square\right) \phi\right],
\end{eqnarray}
\end{widetext}
where ${R}_{\mu \nu}$ is the curvature tensor in the metric formalism.
The scalar curvature $\hat{R}$ can be expressed in terms of the Einstein-Hilbert one as
\begin{eqnarray}
\nonumber \hat{R}&=&g^{\mu \nu}\hat{R}_{\mu \nu}\\
&=&R-\frac{6\omega}{\phi}\square \phi.
\end{eqnarray}
\par Now, varying the action Eq.$\eqref{eq1}$ with respect to the metric tensor leads to 
\begin{widetext}
\begin{eqnarray}
\nonumber	(M_p^2 +\xi \phi^2 )G_{\mu \nu}=(1+2\xi-4\xi \omega) \nabla_{\mu}\phi \nabla_{\nu} \phi
	-\left( \frac{1}{2}+2\xi-\xi \omega\right)g_{\mu \nu } (\nabla \phi)^2 -g_{\mu \nu }V(\phi)+2\xi(1+\omega) \phi\left[  \nabla_{\mu} \nabla_{\nu}-g_{\mu \nu}\square \right] \phi,\\
	\label{eq2.9}
\end{eqnarray}
\end{widetext}
which can be rewritten as
\begin{equation}
    F(\phi)G_{\mu \nu}=\kappa^2T_{\mu \nu},
\end{equation}
where F denotes a function of $\phi$, given by
\begin{equation}
    F(\phi)=1+\xi\kappa^2\phi^2,
\end{equation}
and $ T _{\mu \nu}$ is the matter energy-momentum tensor which takes the form
\begin{eqnarray}
\nonumber T _{\mu \nu}&&= A\nabla_{\mu}\phi\nabla_{\nu}\phi-B g_{\mu \nu}(\nabla \phi)^2-g_{\mu \nu}V(\phi)\\
 &&+C\phi\left[ \nabla_{\mu}\nabla_{\nu}-g_{\mu \nu}\square \right] \phi,
\end{eqnarray}
with $A=(1+2\xi-4\xi \omega)$, $B=\left( \frac{1}{2}+2\xi-\xi \omega\right)$, and \\
$C=2\xi(1+\omega)$ are constants.\\

\par In the case of $\omega=0$, Eq \eqref{eq2.9} describes NMC in  the metric approach \cite{Komatsu:1997hv}. Meanwhile, in the case $\xi=0$, we recover the case of general relativity.

\par Finally, let us take the variation of the action Eq.$\eqref{eq1}$ with respect to $\phi$, to get the modified Klein Gordon equation \cite{24}
\begin{equation}
 \square \phi  + \xi \hat{R}\phi - V_{,\phi}=0,
\label{eq2.12}
\end{equation}
where $\square \phi = \frac{1}{\sqrt{-g}}\partial_\nu (\sqrt{-g}g^{\mu \nu} \partial _{\mu} \phi)$ is the D'Alembertien and  $V_{,\phi}=dV/d\phi$.
\section{SLOW ROLL EQUATIONS}
\label{secIII}
In this section, we assume a homogeneous and isotropic
Universe described by a spatially flat Robertson-Walker (RW) metric  with the signature (-,+,+,+) \cite{25}
\begin{eqnarray}
ds^2=-dt^2+a^2(t)(dx^2+dy^2+dz^2),
\end{eqnarray}
where $a(t)$ is the scale factor and t is the cosmic time.
The Friedmann equation is acquired by taking the 00 component of Eq.$\eqref{eq2.9}$
\begin{eqnarray}
\nonumber H^2=\frac{\kappa^2}{3F(\phi)}\left[ \left( \frac{1}{2}-3\xi\omega\right) {\dot{\phi}}^2+V(\phi)-6H\xi(1+\omega) \phi \dot{\phi}\right], \\
\label{eq 2.3}
\end{eqnarray}
where $H=\dot{a}/a$ is the Hubble parameter, and a dot denotes the differentiation with respect to cosmic time. In the slow roll conditions $\frac{\dot{\phi}}{\phi}<< H$ and $\dot{\phi}^2<<V$, Eq.$\eqref{eq 2.3}$ can be approximated by
\begin{equation}
H^2	\simeq\frac{\kappa^2 V(\phi)}{3(1 +\xi\kappa^2 \phi^2 )}.
\end{equation}
\par By replacing $\square \phi$, $\hat{R}$ and $R$ by their expressions, the inflaton field equation Eq.$\ref{eq2.12}$ becomes
\begin{equation}
-3H\dot{\phi}(1-6\xi\omega)+12\xi\phi H^2 -V_{,\phi}\simeq0.
\label{eq3.4}
\end{equation} \\
\section{SCALAR PERTURBATIONS}
\label{secIV}
In this section, we present in detail the scalar cosmological perturbations. We choose the Newtonian gauge, in which the scalar metric perturbations of a RW background are given by \cite{26,27}
\begin{eqnarray}
	ds^2=-(1+2\Phi)dt^2+a(t)^2(1-2\Psi)\delta_{ij}dx^idx^j, \label{4.1}
\end{eqnarray}
where $\Phi(t,x)$ and $\Psi(t,x)$ are the scalar perturbations called also Bardeen variables \cite{28}.
\par  The perturbed Einstein's equations are given by
\begin{eqnarray}
  	\delta F(\phi)G^\mu_{\nu}+ F(\phi)\delta G^\mu_{\nu}=\kappa^2\delta T^\mu_{\nu}.
  	\label{eq4.2}
\end{eqnarray}
\par For the perturbed metric Eq.$\eqref{4.1}$, we obtain the individual components of Eq. $\eqref{eq4.2}$ in the following form
\begin{widetext}
\begin{subequations}
\begin{equation}
-6\xi \kappa^2H^2\phi\delta\phi+F(\phi)\left[ 6 H(\dot{\Psi}+H\Phi)-2\frac{\nabla^2}{a^2}\Psi\right]
=\kappa^2\delta T^{0}_{0},
\end{equation}
\begin{equation}
-2F(\phi)(\dot\Psi+H\Phi)_{,i}=\kappa^2\delta T^{0}_{i},\\
\end{equation}

\begin{equation}
 -6\xi \kappa^2\phi\delta\phi(3H^2+2\dot H)
 +6F(\phi)\left[ (3H^2+2\dot H)\Phi+H(\dot\Phi+3\dot\Psi)+\ddot\Psi+\frac{\nabla^2}{3a^2}(\Phi-\Psi)\right] 
=\kappa^2\delta T^{i}_{i}, 
\end{equation}

\begin{equation}
F(\phi)a^{-2}(\Psi-\Phi)^{,i}_{,j}=\kappa^2\delta T^{i}_{j}.
\end{equation}
\label{eq4.3}
\end{subequations}
\end{widetext}

The perturbed energy momentum tensor $\delta T^\mu_\nu$ appearing in Eq.$\ref{eq4.2}$ is given by \cite{29}
\begin{eqnarray}
\delta T^\mu _\nu=\begin{pmatrix} -\delta \rho  & a\delta q_{,i} \\ -a^{-1}\delta q^{,i} & \delta p \delta^i_j+\delta \pi ^i_j \end{pmatrix},
\end{eqnarray}
where $\delta \rho$, $\delta q$ and  $\delta p$ represent the perturbed energy density, momentum, pressure, respectively. The anisotropic stress tensor is given by $\delta \pi^i _{j}=\left(   \triangle^i_{j}-\frac{1}{3}\delta^i_{j}  \triangle\right) \delta \pi$ where $\triangle^i_{j}$ is defined by $\triangle^i_{j}=\delta^i_k\partial_k\partial_j$ and $\triangle=\triangle^i_{i}$.\\

\par Now, let us simplify the calculation and study the evolution of perturbations. To do so, we decompose the function $\psi(x,t)$ into its Fourier components
$\psi_k(t)$ as follows
\begin{eqnarray}
\psi (t,x)=\frac{1}{(2\pi)^{3/2}} \int e^{-ikx}\psi_k (t)d^3k,
\end{eqnarray}
where $k$ is the wave number. The perturbed equations Eq.$\eqref{eq4.3}$ can be expressed as
\begin{widetext}

\begin{subequations}
\begin{equation}
-\xi \kappa^2H\phi\delta\phi+F(\phi)\left[ H(\dot{\Psi}+H\Phi)+\frac{k^2}{3a^2}\Psi\right] =\frac{-\kappa^2}{6}\delta \rho,\\
\end{equation}
\begin{equation}
F(\phi)(\dot\Psi+H\Phi)=\frac{-\kappa^2}{2}a \delta q,\\
\label{eq4.6b}
\end{equation}

\begin{equation}
-\xi \kappa^2\phi\delta\phi(3H^2+2\dot H)+F(\phi)\left[ (3H^2+2\dot H)\Phi+H(\dot\Phi+3\dot\Psi)+\ddot\Psi-\frac{k^2}{3a^2}(\Phi-\Psi)\right]=\frac{\kappa^2}{2}\delta p,\\
\end{equation}
\begin{equation}
F(\phi)(\Psi-\Phi)^{,i}_{,j}=\kappa^2 a^{2}\delta \pi^i_{j}.
\end{equation}

\end{subequations}
\end{widetext}

\par By using the perturbed energy momentum tensor, one can write the perturbed energy density, the perturbed momentum, the perturbed pressure, and the anisotropic stress tensor, respectively, as follows:
\begin{widetext}
\begin{subequations}

\begin{align}
 -\delta \rho=2(A-B)\Phi\dot\phi^2-2(A-B)\dot\phi\delta\dot\phi-V_{,\phi}\delta\phi
+3CH\left[ \dot\phi\delta\phi+\phi\delta\dot\phi\right]+6CH(\Psi+\Phi)\phi\dot{\phi}
-C\phi a^{-2}\triangle\delta\phi,
\end{align}
\begin{equation}
    a\delta q =-A\dot{\phi}\delta\phi-C\phi\left( \delta\dot{\phi}-\Phi\dot{\phi}-H\delta\phi\right) ,
    \label{eq4.7b}
\end{equation}

\begin{align}
\delta p=2B(\dot{\phi}
\delta\dot{\phi}-\Phi\dot{\phi}^2)-V_{\phi}\delta\phi+2CH\dot{\phi}\delta\phi
+C\phi\left[2H\delta \dot{\Phi}-2\Phi\ddot{\phi} +\delta\ddot{\phi}-4H\Phi\dot{\phi}-2\dot{\Psi}\dot{\phi} - a^{-2} \triangle\delta\phi\right],
\end{align}
\begin{equation}
   \delta \pi ^i_j=a^{-2}C\phi\delta \phi^{,i}_{,j}.
\end{equation}
\end{subequations}
\end{widetext}
\par The perturbed equation of motion for $\phi$ takes the form
\begin{widetext}
\begin{eqnarray}
\nonumber &2(A-B)\dot{\phi}\delta\ddot{\phi}\\
\nonumber &+\left[ 2(A-B)\ddot{\phi}+V_{\phi}-3C\dot{H}\phi
+6(A-C)H\dot{\phi}-3CH^2\phi\right] \delta\dot{\phi}+\left[ V_{\phi \phi}\dot{\phi}+(A-C)\dot{\phi}\frac{k^2}{a^2}-3CH^2\dot{\phi}-2CH\phi \frac{k^2}{a^2}\right]\delta\phi & \\
\nonumber &=2(A-B)\left[ \dot{\Phi}\dot{\phi}^2+2\Phi\dot{\phi}\ddot{\phi}\right]+6C\dot{H}\phi(\Phi+\Psi)\dot{\phi}+6CH\left[(\dot{\Phi}+\dot{\Psi})\phi \dot{\phi} + (\Phi+\Psi)(\dot{\phi}^2+\phi \ddot{\phi})\right]+C\phi\dot{\phi}\frac{k^2}{a^2}\Phi &\\
&+6AH\Phi\dot{\phi}^2+30CH^2\Phi\dot{\phi}\phi+18CH^2\Psi\dot{\phi}\phi+6C\phi H (\Phi+\Psi)\ddot{\phi}.& \label{eq4.8}
\end{eqnarray}
\end{widetext}
Therefore, if we adopt the slow roll conditions at large scales $k << aH$, we can neglect  $\dot{\Phi}$, $\dot{\Psi}$, $\ddot{\Phi}$ and $\ddot{\Psi}$ \cite{amendola2006constraints, amendola2007solar}.  In fact, throughout the cosmic history of the Universe, significant scales have primarily existed well beyond the Hubble radius and they have only recently reentered in the Universe. Consequently, it is reasonable to consider large scales as a valid assumption. Indeed, to be able to satisfy the longitudinal post-Newtonian limit, we need to consider that $ \mathop{{}\Delta} \Phi\gg a^2H^2\times(\Phi,\dot{\Phi},\ddot{\Phi})$ and similarly for the other gradient terms. In the case of plane wave perturbation with wavelength $\lambda$, when the condition $\lambda\ll 1/H$ is met, we notice that $H^2\Phi$ is much smaller than $\mathop{{}\Delta} \Phi$. For $\dot{\Phi}$ to be also negligible, the condition $\frac{d log \Phi}{d log a}\ll\frac{1}{\lambda H^2}$ is required, and this is satisfied if $\lambda\ll 1/H$ for perturbation growth. The same arguments may be used for $\ddot{\Phi}$ and also for the metric potential $\Psi$ \cite{amendola2006constraints,nozari2012braneworld}. Hence, we can rewrite Eq.$\eqref{eq4.8}$ as follows

\begin{widetext}
\begin{eqnarray}
   \nonumber &(1-6\xi\omega)\delta\ddot{\phi}+\left[ \frac{V_{,\phi}}{\dot{\phi}} +6(1-6\xi\omega)H-6\xi(1+\omega) H^2\frac{\phi}{\dot{\phi}}\right] \delta\dot{\phi}+\left[ V_{,\phi \phi}-6\xi(1+\omega) H^2\right] \delta {\phi}&\\
   &+6H\left[(1+4\xi-2\xi\omega)\dot{\phi}+10\xi(1+\omega)H \phi \right] \Phi=0.&
\label{eq4.9}
\end{eqnarray}
\end{widetext}

Using Eq.$\eqref{eq4.6b}$ and Eq.$\eqref{eq4.7b}$, the scalar perturbation $\Phi$ can be expressed in terms of the fluctuation of the scalar field $\delta\phi$ as
 \begin{eqnarray}
 \Phi=\frac{\kappa^2_{eff}\left( A\dot{\phi}-CH\phi \right) }{2F(\phi)H}\delta \phi,
 \end{eqnarray}
 where $\kappa^2_{eff}=\kappa^2/\left[ 1+\frac{C\kappa^2}{2F(\phi)H}\phi \dot{\phi}\right]$.\\
\par We define the comoving curvature perturbation as follow: \cite{30}
\begin{eqnarray}
 R=\Psi-\frac{H}{\rho + p} a\delta q.
 \end{eqnarray}
Hence, by considering the slow roll approximations at large scale, and from Eq.$\eqref{eq4.6b}$, one can find
 \begin{eqnarray}
 R=\Psi+\frac{H}{\dot{\phi} \left[ 1+\frac{C\kappa^2}{2F(\phi)H}\phi \dot{\phi}\right]} \delta\phi.
 \label{eq4.12}
 \end{eqnarray}
\par Considering the spatially flat gauge where $\Psi=0$, and from Eq.$\eqref{eq4.12}$, one define a new variable as follow
 \begin{eqnarray}
\delta \phi_{\Psi}=\delta \phi +\frac{\dot{\phi}}{H}\left[ 1+\frac{C\kappa^2}{2F(\phi)H}\phi \dot{\phi}\right]\Psi.
\end{eqnarray}
Using Eq.$\eqref{eq4.6b}$, in this gauge, one can rewrite Eq.$\eqref{eq4.9}$ as
\begin{widetext}

\begin{eqnarray}
\nonumber&(1-6\xi\omega)\delta\ddot{\phi_\Psi}+3H\left[(1-6\xi\omega)-2\xi H\frac{\phi}{\dot{\phi}}(\omega-2)\right] \delta\dot{\phi_\Psi}&\\
&+ \left[  V_{,\phi \phi}-6\xi\omega H^2-6\kappa^2_{eff} \left((1+2\xi-4\xi\omega)\dot{\phi}-2\xi(1+\omega) H\phi\right)\frac{(1+4\xi-2\xi\omega)\dot{\phi}+10\xi(1+\omega)H \phi }{2 F(\phi)} \right]\delta \phi_\Psi =0.&
\label{eq4.14}
\end{eqnarray}
\end{widetext}
Introducing of the Mukhanov-Sasaki variable  $v=a\delta \phi_\Psi$, allows us to rewrite the perturbed equation of motion Eq.$\eqref{eq4.14}$ as
\begin{equation}
  v''-\frac{1}{\tau^2}\left[ \nu^2-\frac{1}{4}\right]v=0,
\label{eq4.15}
\end{equation}
where the derivative with respect to the conformal time $\tau$ is denoted by the prime, and the term $\nu$ is 
\begin{equation}
  \nu=\frac{3}{2}+\epsilon-\tilde{\eta}+\frac{\tilde{\zeta}}{3}+2\tilde{\chi},
\end{equation}
where we have used the slow roll parametres given by 
  \begin{eqnarray}
 &\epsilon&=1-\frac{\mathcal{H}'}{\mathcal{H}^2}=\frac{1}{2\kappa^2}\left( \frac{V_{\phi}}{V}\right)^2 C_{1}, \label{eq4.17}\\
  &\eta&=\frac{a^2 V _{\phi\phi}}{3\mathcal{H}^2},\\
 &\zeta&=6\xi\omega,\\
 \nonumber &\chi&=\kappa^2_{eff}\left(  (1+2\xi-4\xi\omega)\phi'-2\xi(1+\omega)\mathcal{H}\phi\right)\\ &&\frac{(1+4\xi-2\xi\omega)\phi'+10\xi(1+\omega)\mathcal{H}\phi}{2 F\mathcal{H}^2},
 \end{eqnarray}
 and
  \begin{eqnarray}
  &&\tilde{\eta}=\frac{1}{(1-6\xi\omega)}\eta,\\
  &&\tilde{\zeta}=\frac{1}{(1-6\xi\omega)}\zeta,\\
  &&\tilde{\chi}=\frac{1}{(1-6\xi\omega)}\chi.
  \end{eqnarray}
\par We have also introduced the correction term to the standard expression as 
\begin{equation}
      C_{1}=\frac{F(\phi)}{(1-6\xi\omega)}\left( 1-\frac{4\xi\kappa^2\phi}{F(\phi)}\frac{V}{V_{\phi}}\right) \left( 1-\frac{2\xi\kappa^2\phi}{F(\phi)}\frac{V}{V_{\phi}}\right).
 \end{equation}
This term characterizes the effect of NMC (through the constant $\xi$) and the Palatini approach (through $\omega$).
 
\par  The solution to Eq.$\eqref{eq4.15}$ is given by \cite{31}
  \begin{eqnarray}
  v=\frac{aH}{\sqrt{2k^3}}\left( \frac{k}{aH}\right)^{3/2-\nu}.
  \end{eqnarray}
  \par The power spectrum for the scalar field perturbations reads as \cite{30}
  
  \begin{eqnarray}
  P_{\delta\phi}=\frac{4\pi k^3}{(2\pi)^3}\displaystyle\left\lvert \frac{v}{a} \right\rvert ^2,
   \end{eqnarray}
\par the spectral index of the power spectrum is given by \cite{30}
     \begin{eqnarray}
   n_s-1=\frac{dLnP_{\delta\phi}}{dLnk}\rvert_{k=aH}=3-2\nu,
    \end{eqnarray}

  which can be expressed in terms of slow roll parametres as
\begin{eqnarray}
  n_s=1-2\epsilon+2\tilde{\eta}-\frac{2\tilde{\zeta}}{3}-4\tilde{\chi}.\label{eq4.28}
   \end{eqnarray}
   
The power spectrum of the curvature perturbations is defined as \cite{30}
\begin{eqnarray}
   A^2_s&=&\frac{4}{25}P_R=\frac{4}{25}\frac{4\pi k^3}{(2\pi)^3}\left\lvert R \right\rvert ^2\\
   &=&\left( \frac{2H}{5\dot{\phi}\left[ 1+\frac{C\kappa^2}{2F(\phi)H}\dot{\phi}\phi\right] }\right)^2P_{\delta\phi},
 \end{eqnarray}
assuming the slow-roll conditions, it becomes 
 \begin{eqnarray}
  \nonumber  A^2_s&=&\frac{4}{25(2\pi)^2}\frac{H^4}{\dot{\phi}^2\left[ 1+\frac{C\kappa^2}{2F(\phi)H}\dot{\phi}\phi\right]^2}\\
    &=&\frac{\kappa^6 V^3}{75\pi^2V^2_{,\phi}}C_{2},
    \label{eq4.31}
  \end{eqnarray}
where 
\begin{eqnarray}
    C_{2}=\frac{(1-6\xi\omega)^2}{F(\phi)\left[ 1+\frac{C\kappa^2}{2F(\phi)H}\dot{\phi}\phi\right]^2}\frac{V^2_\phi}{(2F_{,\phi}V-FV_{,\phi})^2},
\end{eqnarray}
is a correction to the standard expression of the power spectrum.  This correction term depends on NMC and the Palatini approach effect.
\section{TENSOR PERTURBATIONS} \label{secV}
The tensor to scalar ratio is one of the important observable parameter in cosmology. Observational data \cite{Akrami:2018odb} gives an upper limit on this ratio, $r < 0.1$, at a 95\% confidence level. To introduce this parameter, we need to define the tensor perturbations amplitude as \cite{Lidsey:1995np}
\begin{eqnarray}
A^2_T=\frac{2\kappa^2}{25}(\frac{H}{2\pi})^2,
\end{eqnarray}
which, in our model, takes the following form
\begin{eqnarray}
 A^2_T=\frac{4\kappa^4 V}{600\pi^2}C_{3}, \label{eq5.3}
\end{eqnarray}
where the correction term $C_{3}$ is defined as
\begin{equation}
    C_{3}=\frac{1}{F(\phi)}.
\end{equation}
\par Furthermore, we define the tensor to scalar ratio, which is a very useful inflationary parameter  
\begin{eqnarray}
\nonumber r&=&\dfrac{A^2_T}{A^2_S}\\
&=&\frac{1 }{2\kappa^2}\frac{V^2_\phi}{ V^2} \frac{\left[ 1+\frac{C\kappa^2}{2F(\phi)H}\dot{\phi}\phi\right]^2}{(1-6\xi\omega)^2}. \label{eq5.6}
\end{eqnarray}
\\
\section{HIGGS INFLATION}\label{secVI}
In this section, as an application, we study a Higgs inflationary model, in which we consider that the Higgs boson (the inflaton) is NMC to the gravity, within the hybrid metric Palatini approach developed in the previous sections. We will also check the viability of the model by comparing our results with the observational data \cite{Akrami:2018odb}. In this case, we consider the quartic potential \cite{tenkanen2019minimal}
\begin{equation}
V(\phi)=\frac{\lambda}{4}\phi^4,
\end{equation} 
where $\lambda$ is the Higgs self-coupling.
During inflation, the number of e-folds is given by \cite{nozari2013some}
\begin{equation}
 N=\int_{t_I}^{t_F}Hdt=\int_{\phi(t_I)}^{\phi(t_F)}\frac{H}{\dot{\phi}}d\phi.
\end{equation}
From Eq.$\eqref{eq3.4}$
\begin{equation}
\dot{\phi}=\frac{12\xi\phi H^2-V_\phi}{3H(1-6\xi\omega)},
\end{equation}
we get
 \begin{eqnarray}
N=\frac{(1-6\xi\omega)\kappa^2}{8}\left[ \phi^2(t_I)-\phi^2(t_F)\right], \label{eq.4.6}
\end{eqnarray}
where the subscript I and F represent the crossing horizon and the end of inflation, respectively. Considering $\phi^2(t_I)>>\phi^2(t_F)$, we get 
\begin{equation}
\phi^2(t_I)=\frac{8N}{\kappa^2(1-6\xi \omega)}
\end{equation}
\par
\begin{figure}
  \centering
  \includegraphics[width=0.45\textwidth]{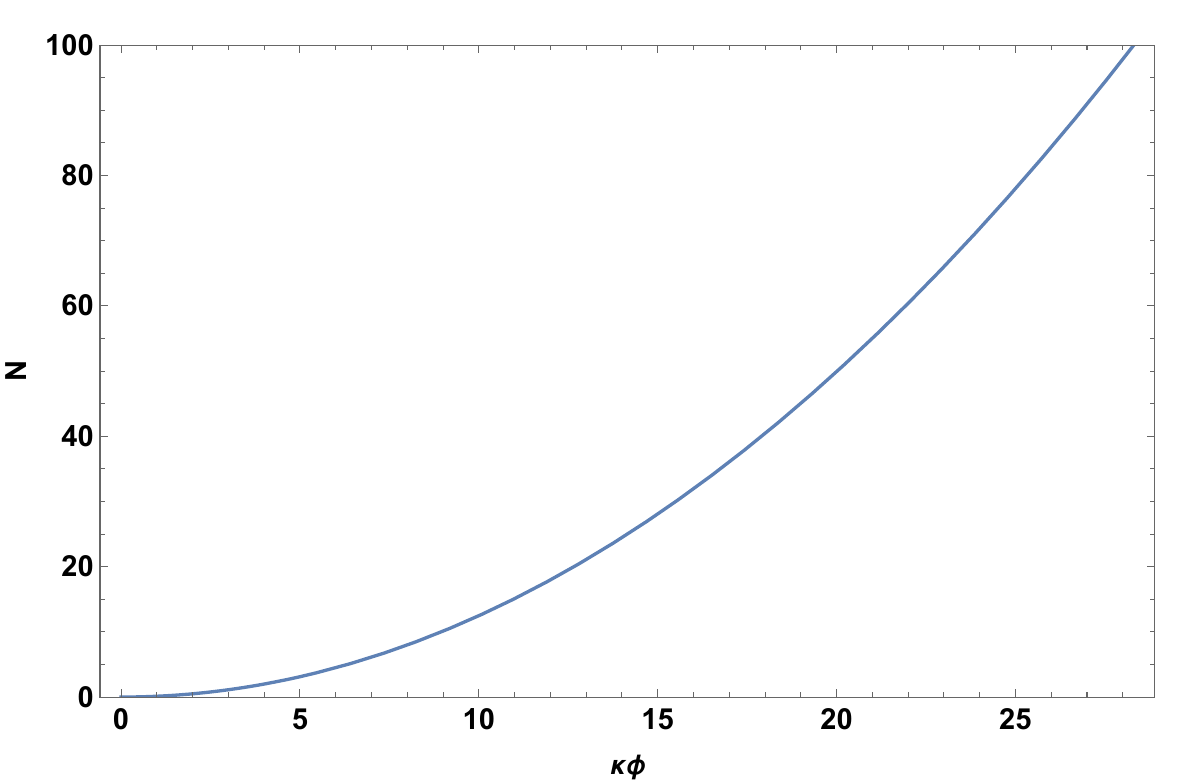} 
\caption{Plot of the number of e-folds versus the scalar
field $\phi$ for $\xi=10^{-3.5}$ and $\lambda=0.13$.}
   \label{fig:1}
 \end{figure}

  In Fig. \ref{fig:1}, We show the variation of the number of e-folds, N, versus the scalar field for a Higgs self-coupling $\lambda=0.13$ \cite{bargach2020nonminimal} and a coupling constant $\xi=10^{-3.5}$. From this figure, we notice that for an appropriate range of N, i.e. $50<N<70$, we get a large field  where $\kappa\phi\gg20$.                                                                                                 
\captionsetup[figure]{justification=Justified}
\par The slow roll parameter defined in Eq.$\eqref{eq4.17}$ becomes 
\begin{equation}
\epsilon=\frac{8}{\kappa^2 \phi^2 (1-6\xi \omega)}(1-\frac{\xi \kappa^2\phi^2}{2F}),
\end{equation}
\begin{figure}[!ht]
\centering
  \includegraphics[width=0.85\linewidth]{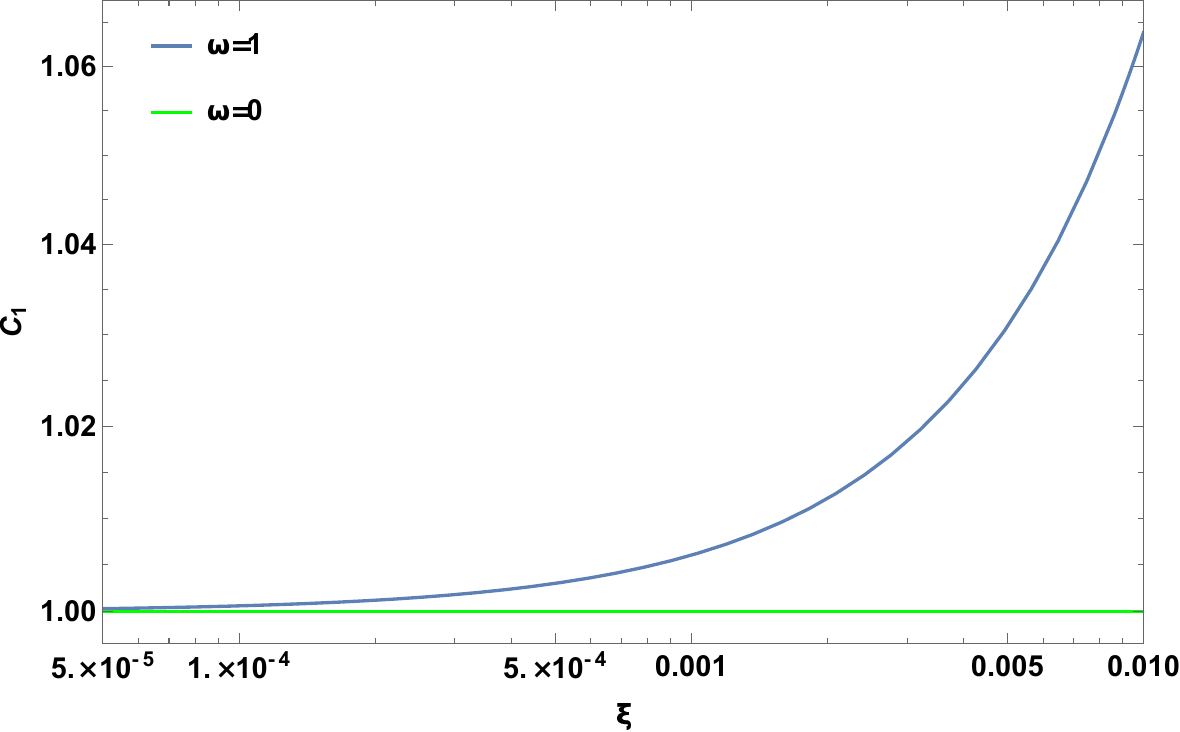} 
\caption{Variation of the correction term $C_1$ as a function of the coupling constant for $N=45$.} 
 \label{fig:2}
\end{figure}

 \par Fig. \ref{fig:2} represents the evolution of correction term $C_1$ as a function of the coupling constant $\xi$. As we can see, the effect of the Palatini parameter $\omega$ on $C_1$ begin from almost the value $10^{-4}$. We can see also that for $\xi=0$, the correction term reduces to one and the standard expression of the slow roll parameter is recovered. While, for $\xi\neq0$ and $\omega=0$ we recover the slow roll parameter expression in the case of NMC within the metric approach.\\
\par The spectral index of the power spectrum  given by Eq.$\eqref{eq4.28}$, can be written as follow
\begin{widetext}
\begin{eqnarray}
\nonumber n_s&=&1- \frac{16}{\kappa^2 \phi^2(1 - 6 \xi \omega)} (1 - \frac{\xi \kappa^2  \phi^2}{2F})\\
 &+&\frac{2}{(1 - 6 \xi \omega) } [\frac{12 F}{\kappa^2  \phi^2} - 2 \xi \omega - \kappa_{eff} ((1 - 4 \xi \omega + 2 \xi) \dot{\phi} - 2 \xi(1+\omega) H \phi)\frac{ (1 - 2 \xi \omega +4\xi ) H \dot{\phi} +10 \xi(1+ \omega) H^2 \phi}{F H^3}].
\end{eqnarray}
\end{widetext}
\par
\begin{figure*}
\centering
\begin{subfigure}[!ht]{0.45\linewidth}
\centering
\includegraphics[width=\linewidth]{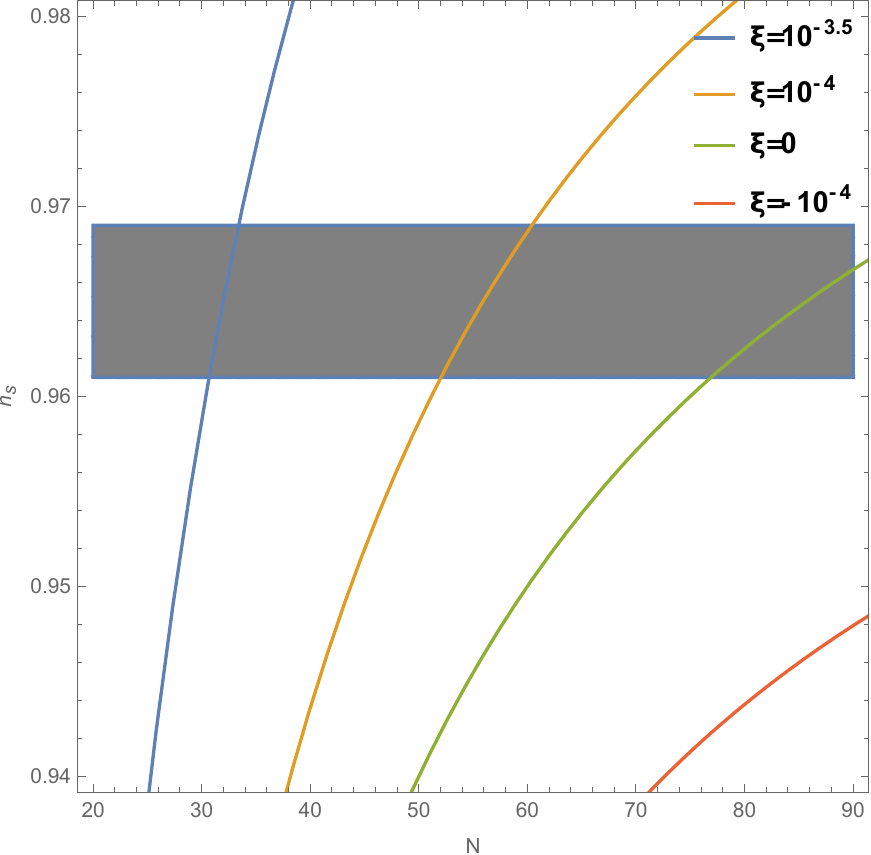} 
     \caption{} 
\end{subfigure}
\hfill
\begin{subfigure}[!ht]{0.45\linewidth}
\centering
\includegraphics[width=\linewidth]{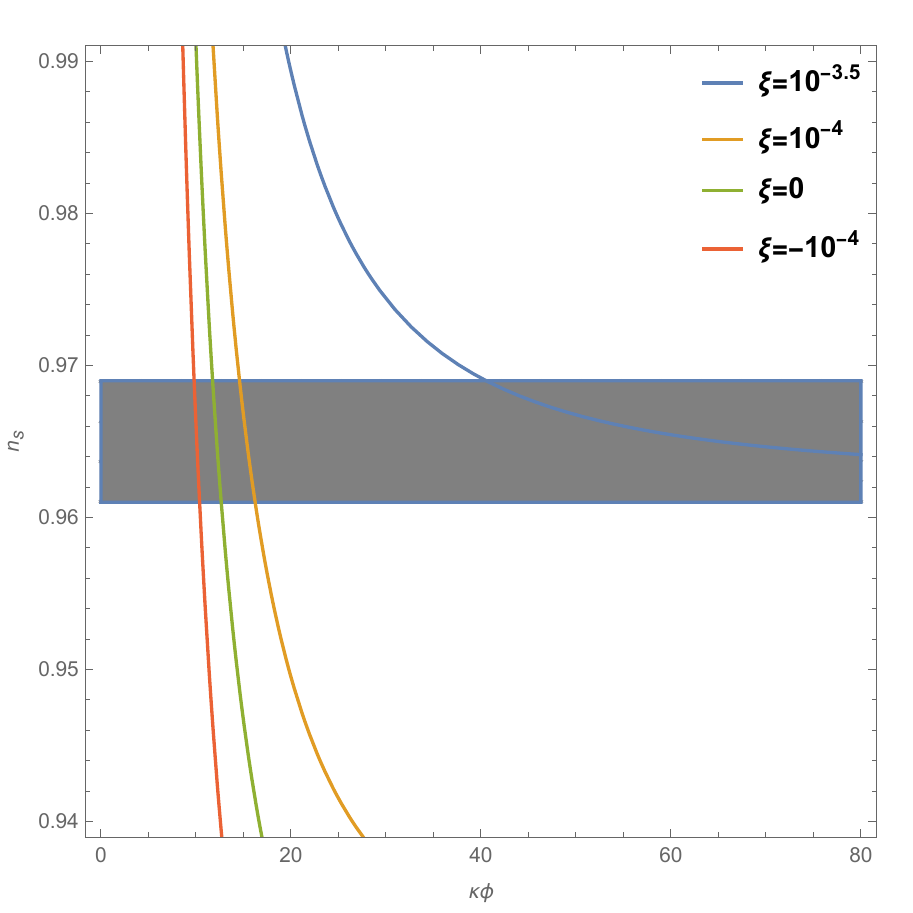} 
  \caption{}
\end{subfigure}
\caption{Evolution of $n_s$ against the number of e-folds (a) and against the scalar field (b) for different values of the coupling constant $\xi$ and $\lambda=0.13$.}
\label{fig:3}
\end{figure*}

 Figs. \ref{fig:3}(a) and \ref{fig:3}(b) illustrate the variation of $n_s$ against the number of e-folds $N$ and against the scalar field for $N=45$, respectively, for $\lambda=0.13$ and for different values of the coupling constant $\xi$ i.e. $10^{-3.5}, 10^{-4}, 0,$ and $-10^{-4}$.  The gray horizontal bound in both figures represents the limits for the spectral index imposed by Planck data. We conclude that the predictions of $n_s$ are consistent with the observational data for $\xi=10^{-4}$ and $\xi=10^{-3.5}$.

\par From Eq.\eqref{eq4.31} and Eq.\eqref{eq5.3}, we get the power spectrum of the curvature perturbations and the tensor perturbations amplitude as
 
\begin{eqnarray}
A^2_s=\frac{\lambda\kappa^6 \phi^6}{4800 \pi^2}C_2,
\end{eqnarray}

\begin{eqnarray}
A^2_T=\frac{\lambda\kappa^4\phi^4 }{600\pi^2}C_3,
\end{eqnarray}
respectively.\\
\par
\begin{figure}[!ht]
\centering
\includegraphics[width=0.85\linewidth]{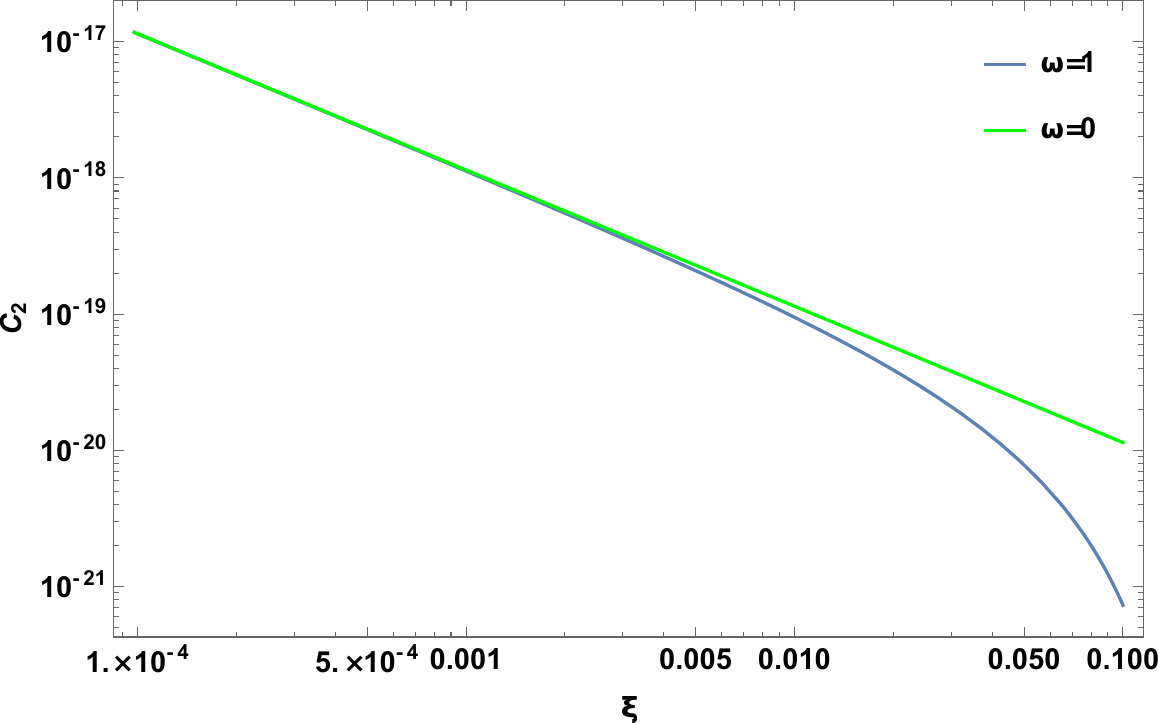} 
\caption{Variation of the correction term $C_2$ versus the coupling constant for the number e-folds $N=45$.} 
	\label{fig:4}
\end{figure}

 The behavior of $C_2$ is shown in Fig. \ref{fig:4}. We present this term versus the coupling constant $\xi$ in the cases of the hybrid Palatini metric formalism (blue curve) and in the metric formalism (green curve). The effect of the Palatini parameter $\omega$ on $A^2_s$ appear from $\xi=5\times10^{-3}$.

\par
\begin{figure}[!ht]
\centering
\includegraphics[width=0.85\linewidth]{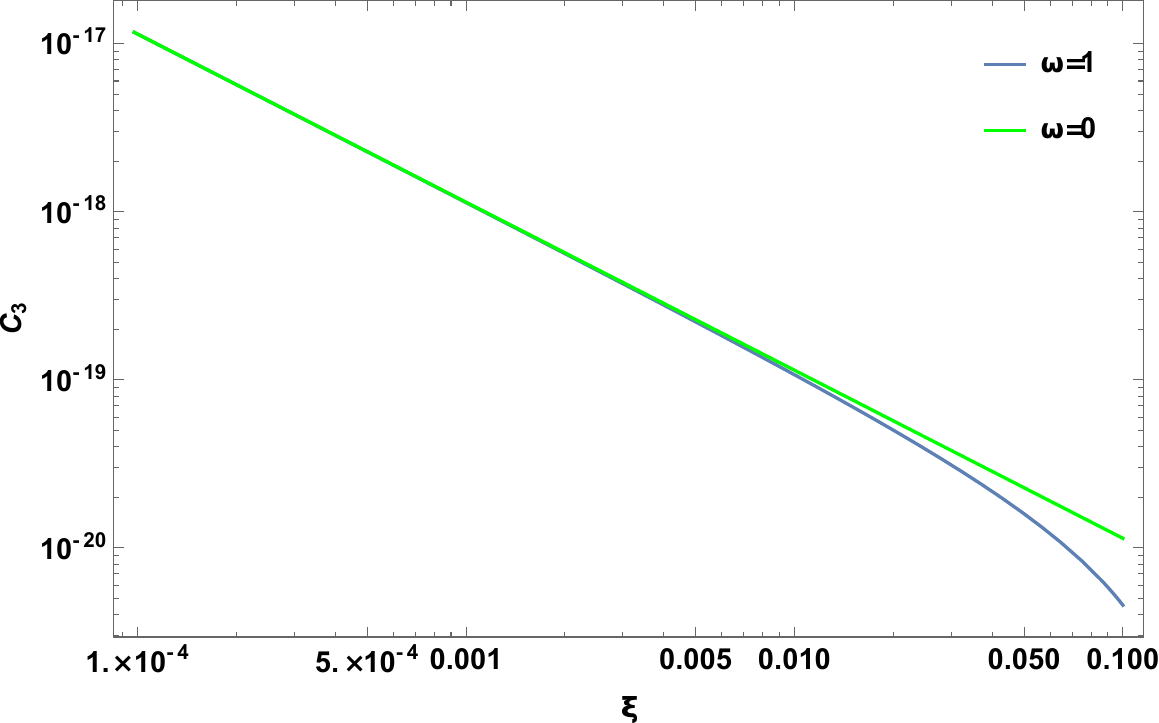} 
\caption{Variation of the correction term $C_3$ against the coupling constant for the number e-folds $N=45$.} 
	\label{fig:5}
\end{figure}

 The correction term, $C_3$, is plotted as a function of $\xi$  in Fig. \ref{fig:5}. We notice that the effect of the Palatini parameter appears from  the value $\xi= 10^{-2}$.
 
From Eq.$\eqref{eq5.6}$, we find that the tensor to scalar ratio can be obtained as
\\
\begin{eqnarray}
r=\frac{8}{\kappa^2\phi^2}\frac{\left[ 1+\frac{ C\kappa^2}{2F(\phi)H}\dot{\phi}\phi\right]^2}{(1-6\xi\omega)^2}.
\end{eqnarray}

\par
\begin{figure}[!ht]
\centering
\includegraphics[scale=0.42]{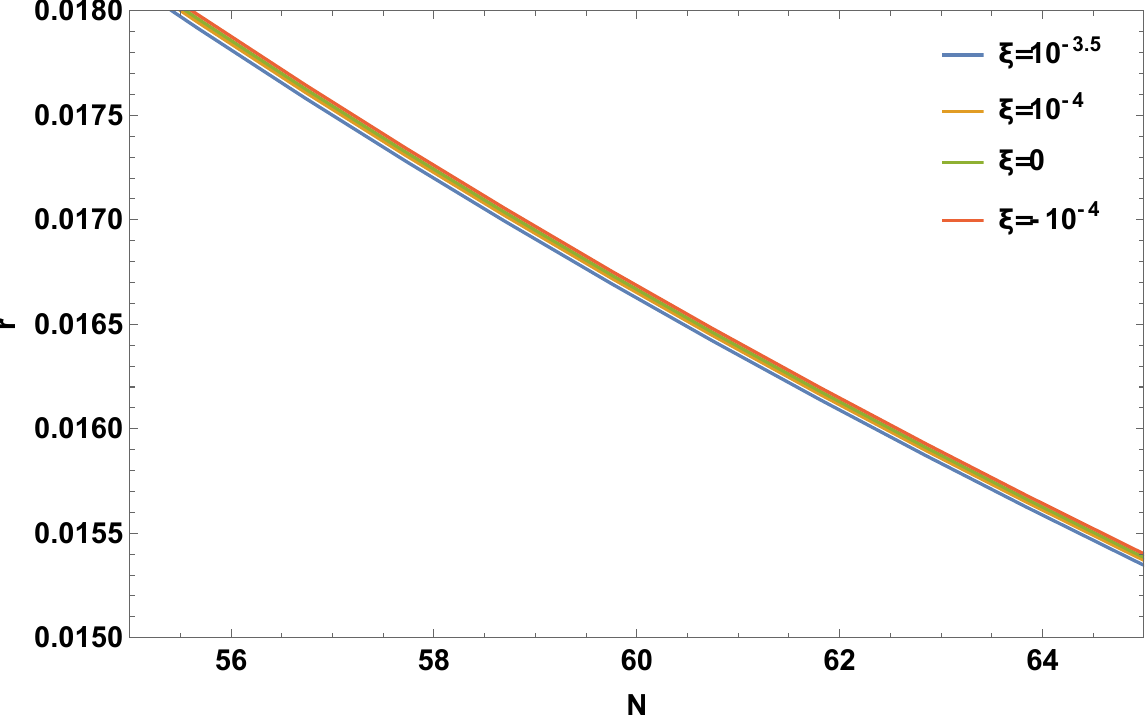} 
\caption{Variation of the tensor to scalar ratio r as a function of the number of e-folds $N$ for different values of the coupling constant.} 
	\label{fig:6}
\end{figure}

 In Fig. $\ref{fig:6}$, we present  the evolution of r versus the number of e-folds N for $\lambda=0.13$ and for selected values of the coupling constant $\xi$. We notice that r lies within the bounds imposed by observational data \cite{Akrami:2018odb} in the appropriate range of N for the selected values of $\xi$.

\begin{figure}[!ht]
\centering
\includegraphics[scale=0.58]{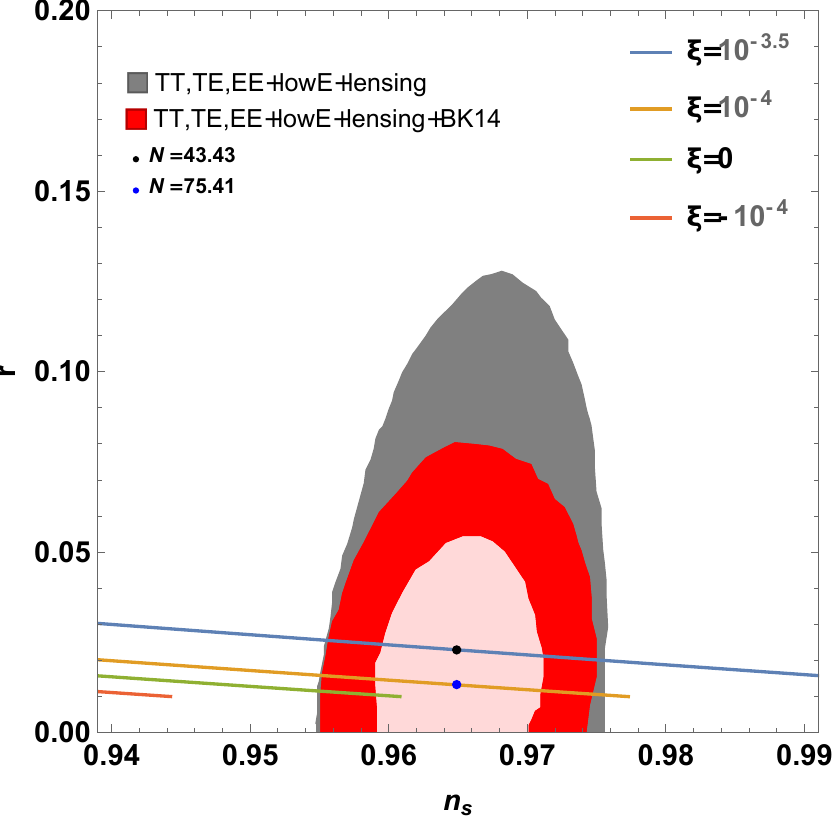} 
\caption{Variation of the tensor to scalar ratio $r$ against the scalar spectral index $n_s$ for selected values of the coupling constant. The gray and the red contour correspond to the Planck TT,TE,EE+LowE+lensing and the Planck TT,TE, EE+lowE+lensing+BK14 data, respectively.}
\label{fig:7}
\end{figure}

\par Fig. \ref{fig:7} shows the $(n_s, r)$ plane for different values of the coupling constant $\xi$ in the range of the number of e-folds $30 \leq N \leq 90$ with the constraints from the Planck TT,TE,EE+LowE+lensing (gray contour) as well as  Planck TT,TE, EE+lowE+lensing+BK14 data (red contour). We notice that $n_s-r$ predictions for the case where $\xi\leq 0$ are ruled out at $95\%$ confidence level contour given by the current observational data \cite{Akrami:2018odb}. Furthermore, for $\xi=10^{-3.5}$, observational parameters lie within $68\%$ CL contour for the range of e-folds number $40.9\leq N\leq 47$ (low-N scenario). In addition, we obtain the central value of the index spectral $n_s=0.9649$ with the small value of tensor to scalar ratio $r=0.022$ for $N=43.43$. On the other hand, for $\xi=10^{-4}$, the results are inside the $68\%$ CL contour for the range $67.8\leq N \leq 86$ (high-N scenario). However, $N=75.41$ give $n_s=0.9649$ and $r=0.013$. Thus, one conclude that NMC in the framework of hybrid metric Palatini can ensure successful Higgs inflation. In the literature, they found that NMC in the Pure Palatini formalism requires large value of $\xi$ and give an extremely small value of tensor to scalar ratio $r\sim 10^{-12}$ \cite{Rasanen:2017ivk, Markkanen:2017tun, Takahashi:2018brt}. Therefore, the hybrid model may be an effective way to solve this issue by increasing the value of $r$ and make it comparable with the corresponding values predicted by the original metric approach and then may be probed by future experiments \cite{Matsumura:2016sri, Kogut:2011xw, Sutin:2018onu}, where the value of the tensor to scalar ratio is of order $r\sim 10^{-2}$.
\section{CONCLUSION}\label{secVII}
In this work, we have studied a cosmological model where the field is non-minimally coupled with gravity in the hybrid metric Palatini approach. \par

We have also analyzed the cosmological perturbations in order to determine the different parameters during the inflationary period. As we have already mentioned, the existence of correction
terms to the standard background and perturbative parameters, represents the impact of the Palatini approach and the  non-minimal coupling between the scalar field and the Ricci scalar. \par

We have applied our model by developing in detail a non-minimally coupled inflationary model driven by the Higgs field with a quartic potential, within the slow-roll approximation. \par 

We have checked our results by plotting the evolution of different inflationary parameters
versus the constraints provided by the observational data as shown in the figures.

We have found that the perturbed parameters such as the tensor to scalar ratio and the scalar spectral index are compatible with the observational data, for an appropriate range of the number of e-folds for different values of $\xi$ as Figs. \ref{fig:3} and \ref{fig:6} show.\par

We have plotted the different correction terms to the standard case versus the coupling constant. We have showed that they are depending on NMC and the Palatini effect.

Finally, for more checking of consistency of our model, we have compared our theoretical predictions with
observational data \cite{Akrami:2018odb} by plotting the Planck confidence contours in the plane of $n_{s}-r$ (Fig. $\ref{fig:7}$). The results show that the predicted parameters are in good agreement with the Planck data for two values of NMC constant $\xi=10^{-3.5}$ and $\xi=10^{-4}$.

\end{document}